\documentclass{webofc}
\usepackage[varg]{txfonts}   
\usepackage{float}
\usepackage{slashed}
\usepackage{bm}
\woctitle{QCD@Work 2014}
\begin{document}
\title{Analysis of $\rho$ condensation in a Nambu-Jona-Lasinio model}
%
%

\author{Marco Frasca\inst{1}\fnsep\thanks{\email{marcofrasca@mclink.it}}
}

\institute{Via Erasmo Gattamelata, 3 \\
           00176 Rome (Italy)
          }

\abstract{%
  A non-local Nambu-Jona-Lasinio model can be derived from QCD in the low-energy limit. In this way, it is possible to fix all the free parameters of the model with physical ones. We used this approach to derive a local limit to the Nambu-Jona-Lasinio model, with the parameters those obtained from QCD, in order to fix the physical parameters of $\rho$ condensation and obtain the critical magnetic field. $\rho$ condensation is a consequence of the highly non-trivial behavior of the QCD vacuum in presence of a very strong magnetic field giving rising to superconductive behavior in quark matter. We extend previous computations by including finite temperature terms.
}
\maketitle
\section{Introduction}
\label{intro}

Recently, Maxim Chernodub put forward the proposal that, at a given critical magnetic field, $\rho$ mesons could form a condensate proper to a superconductive state \cite{Chernodub:2010qx,Chernodub:2011mc,Chernodub:2012tf}. This will characterize a quantum phase transition where it is not the varying of temperature to cause a transition but rather some other control physical parameter that in this case is the magnetic field. Recently, we obtained a non-local Nambu-Jona-Lasinio (NJL) model directly from QCD \cite{Frasca:2011bd,Frasca:2012eq} and so, we were able to characterize this transition obtaining the critical field at zero temperature in \cite{Frasca:2013kka}. NJL model was derived assuming a trivial infrared fixed point for the Yang-Mills theory \cite{Frasca:2010ce} in absence of quarks. The value of the critical field is seen to be rather varying across different authors ranging from $0.6\ GeV^2$ in \cite{Chernodub:2010qx} from a hadronic model to $eB_c>1.0\ GeV^2$ in \cite{Chernodub:2011mc} from a NJL calculation and finally, to lower values in our computation $0.98 M_q^2$, being $M_q\approx 0.214\ GeV$ the effective quark mass \cite{Frasca:2012eq} and $0.2\ GeV^2$ in the most recent work \cite{Liu:2014uwa}. Practically, the square root of our critical field is just a factor 2 away from the same value of this latter group. This points toward a significantly lower value for the critical field of the transition.

In this work, we extend our analysis to finite temperature and yield a phase diagram of the transition. This computation has never been attempted before and the result is rather interesting showing a saturation at increasing magnetic field possibly indicating that the full non-local NJL model model is needed at increasing field and temperature.

\section{QCD in the infrared limit}
\label{sec-1}

As discussed in \cite{Frasca:2011bd,Frasca:2012eq}, QCD at low-energies takes the form of a non-local NJL model given by
\begin{eqnarray}
      S_{NJL}&=&\sum_q\int d^4x\bar q(x)\left[i{\slashed\partial}-m_q\right]q(x)
      +\int d^4x'{\cal G}(x-x')\sum_q\sum_{q'}\bar q(x)\frac{\lambda^a}{2}\gamma^\mu\bar q'(x')\frac{\lambda^a}{2}\gamma_\mu q'(x')q(x) \nonumber \\
      &&+\int d^4x\left[\frac{1}{2}(\partial\sigma)^2-\frac{1}{2}m_0^2\sigma^2\right]
\end{eqnarray}
being $q(x)$ the quark fields, $m_q$ the quark masses and $\sigma(x)$ the field of the ground state of the tower of excitations of the gluon field. The kernel ${\cal G}(x-x')$ determining the non-local NJL model is obtained from the trivial infrared fixed point of the Yang-Mills theory in absence of quarks and its Fourier transform is
\begin{equation} 
\label{eq:G}
   {\cal G}(p)=-\frac{1}{2}g^2\sum_{n=0}^\infty\frac{B_n}{p^2-(2n+1)^2m_0^2+i\epsilon}=\frac{G}{2}{\cal C}(p)
\end{equation}
being $B_n=(2n+1)^2\frac{\pi^3}{4K^3(-1)}\frac{e^{-(n+\frac{1}{2})\pi}}{1+e^{-(2n+1)\pi}}$ with $K(-1)\approx 1.3111028777$ an elliptic integral and $m_0$ the mass gap of the theory. One has ${\cal C}(0)=1$ that is the case we consider in this paper. In this way we fix the value of $G$ using the gluon propagator. The form factor (\ref{eq:G}) yields an almost perfect agreement with the case of an instanton liquid for the vacuum \cite{Schafer:1996wv,Frasca:2011bd}.This can be rephrased by saying that NJL model is not confining and to recover confinement higher order corrections are needed \cite{Frasca:2012iv}.

We limit our analysis to the case of two flavors $q=u,d$. We turn on the electromagnetic interaction, apply a Fierz transformation to the quark fields and bosonize. The final result takes the form \cite{Frasca:2013kka}
\begin{eqnarray}
\label{eq:SB}
      S_B&=&\int d^4x\left[\frac{1}{2}(\partial\sigma)^2-\frac{1}{2}m_0^2\sigma^2
   +\sum_{q=u,d}\bar q\left(i{\hat {\cal D}}-m_q\right)q\right. \nonumber \\
   &&\left.
   -\frac{1}{2G}\left({\sigma^2}+{\bm\pi}\cdot{\bm\pi}\right)
   +\frac{1}{G}\left({\bm V}_a\cdot{\bm V}_a+{\bm A}_a\cdot{\bm A}_a\right)\right]
\end{eqnarray}
being $i{\hat {\cal D}} = i\slashed{\cal D} - m_q + {\slashed {\hat V}} + \gamma^5 {\slashed {\hat A}} - (\sigma +  i\gamma^5 {\bm\pi}\cdot{\bm\tau})$ and $\slashed{\cal D}={\slashed\partial} -i\hat e \, {\slashed {\cal A}}$. Here the vector part is composed by the flavor--singlet coordinate--vector $\omega$--meson field $\omega_\mu$, and of the electrically neutral, $\rho^0_\mu \equiv \rho^3_\mu$, and charged, $\rho^\pm_\mu = (\rho^{1}_\mu \mp i \rho^{2}_\mu)/\sqrt{2}$, components of the $\rho$-meson triplet, and four pseudovector (axial) fields. We also note that the mass gap gets a correction $1/G$ for the $\sigma$ field \cite{Frasca:2011bd}. Finally, one chooses ${\cal A}_\mu=\left(0,Bx_2/2,-Bx_1/2,0\right)$ with $B$ a constant magnetic field.

\section{$\rho$ mass at finite $T$ and $B$}

The mass of the $\rho$ from a NJL model is already known in literature and we assume it to be given by \cite{Ebert:1994mf,Bernard:1993rz}
\begin{equation}
    m_\rho^2=\frac{3}{8G'N_cJ_2(p^2=0)}
\end{equation} 
being
\begin{equation}
    J_2(p^2=0)=-iN_f\int\frac{d^4q}{(2\pi)^4}\frac{1}{(q^2-v^2+i\epsilon)^2}
\end{equation}
that is clearly divergent and needs regularization. In order to go at finite temperature and non-zero magnetic field, we use the following changes in the mass formula
\begin{enumerate} 
\item Change euclidean integrals to
  $\int\frac{d^4p_E}{(2\pi)^4}\rightarrow T\sum_q\frac{|e_qB|}{2\pi}\sum_k\beta_k\sum^{\infty}_{n=-\infty}\int\frac{dp_z}{2\pi}$ with $\beta_k=2-\delta_{k,0}$.
\item Change $p_0\rightarrow (2n+1)\pi T$.
\item Change the third component of momenta to
    $p_z^2\rightarrow p_z^2+2k|e_qB|$.
\item Finally, regularize it.
\end{enumerate}
These rules were firstly introduced in \cite{Menezes:2008qt} and adopted in \cite{Frasca:2013kka} for this kind of studies. Then, one has
\begin{equation}
     J_2(B,T)=T\sum_q\frac{|e_qB|}{2\pi}\sum_k\beta_k\sum^{\infty}_{n=-\infty}\int_{-\infty}^{+\infty}\frac{dp_z}{2\pi}
\frac{1}{[(2n+1)^2\pi^2T^2+p_z^2+2k|e_qB|+M_q^2]^2}.
\end{equation}
In this equation we assume for $M_q$ the solution of the gap equation at the leading order for $T$ and $B$ going to 0. In our case this means $M_q\approx 0.214\ GeV$ as in \cite{Frasca:2012eq}. This is another approximation we adopt. We are interested in the limit of decreasing $T$ ($T\rightarrow 0$). Now, we just note that a quantum phase transition could occur when some other parameter is varied rather than the temperature (taken to be 0). In our case the only remaining parameter is the magnetic field. What we have to show is that the integral $J_2$ changes sign crossing a 0, at T=0, for a critical magnetic field. We observe that such a zero in $J_2$ is unphysical and is just signaling that non-local effects come into play at the critical magnetic field, as also pointed out more recently \cite{Chernodub:2014rya}, modifying the vacuum of QCD in a significant way. Indeed, at this critical value of the magnetic field, we cannot anymore accept a contact interaction in the NJL model and the full non-local model should be considered instead.

\begin{figure}[H]
\centering
\includegraphics[width=10cm,clip]{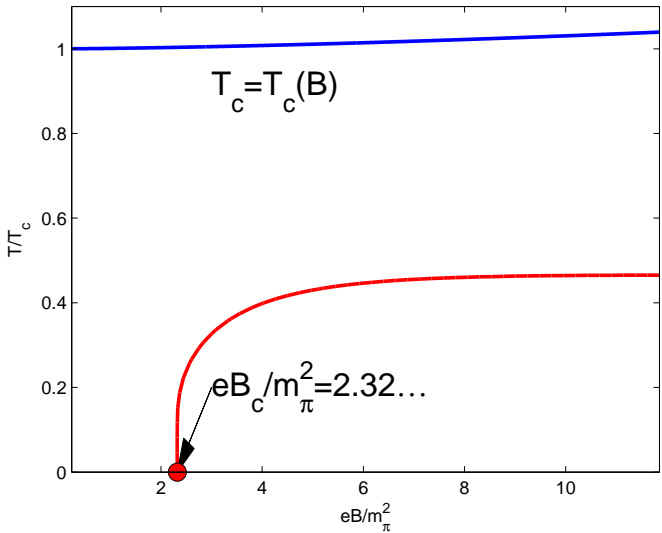}
  \caption{Phase diagram for $\rho$ condensation (the curve $T_c=T_c(B)$ comes from \cite{Gatto:2010pt}).}
\label{fig-1}       
\end{figure}

Evaluating the sum and the integral yields the result
\begin{equation}
   J_2(B,T)\approx\frac{1}{16\pi^2}\sum_q\left[-\frac{|e_qB|}{v^2}-\psi\left(\frac{v^2}{2|e_qB|}\right) 
   -\frac{2|e_qB|}{T}\sum_{k=0}^\infty\frac{F\left(\frac{\sqrt{2k|e_qB|+v^2}}{T}\right)}{\sqrt{2k|e_qB|+v^2}}\right].
\end{equation}
being $F(a) = \pi-2a\left[K_0(a)+\frac{\pi}{2}(K_0(a){\bm L}_1(a)+K_1(a){\bm L}_0(a))\right]$ where $K_n$ are Macdonald functions, ${\bm L}_n$ is the modified Struve functions and $\psi(x)$ is the digamma function. The contributions from higher levels are negligible small. This means that our final formula is
\begin{equation}
   J_2(B,T)\approx\frac{1}{16\pi^2}\sum_q\left[-\frac{|e_qB|}{v^2}-\psi\left(\frac{v^2}{2|e_qB|}\right) 
   -\frac{2|e_qB|}{T}\frac{F\left(\frac{v}{T}\right)}{v}\right].
\end{equation}
This admits a zero solution $B_c=B_c(T)$ with $B_c(0)$ finite, the squared mass of the $\rho$ appears to change sign, and so there is a quantum phase transition at the critical field indicated in Fig.~\ref{fig-1} as also found in \cite{Frasca:2013kka}. It is interesting to note that our $\sqrt{eB_c}$ is just a factor 2 away from the critical field found in \cite{Liu:2014uwa}. This latter computation being more precise than ours, we see a significant lowering of the critical field from NJL model than found in the initial estimations. Besides, the curve of the critical field appears to reach an almost flat behavior as the magnetic field increases. This could indicate that our local approximation is excessive and the full non-local model is needed at increasing magnetic field and temperature.

\section{Conclusions}

We have derived the critical field of the $\rho$ condensation at varying temperature. We have seen as the result persists and the critical field shows a tendency to increase at increasing temperature. We have also seen that the curve appears to saturate at increasing field and we attributed this to the local approximation in the NJL model that is non-local instead. The most interesting aspect is that we obtained a significantly smaller critical field for the transition whose square root is just a factor 2 away from a more precise computation performed by another group recently.

\section*{Acknowledgements}
\thispagestyle{empty}

I would like to acknowledge interesting and useful discussions with Mei~Huang for which I am grateful.

%
%
%

\end{document}